\pdfoutput=1

\documentclass[10pt,twocolumn,prl,aps,amssymb,amsmath,tightenlines]{revtex4}
\usepackage{graphicx}

\newcommand{\cPT}{\ensuremath{\mathcal{PT}}}
\newcommand{\half}{\mbox{$\textstyle{\frac{1}{2}}$}}
\newcommand{\quarter}{\mbox{$\textstyle{\frac{1}{4}}$}}

\begin{document}

\title{Scattering off $\cPT$-symmetric upside-down potentials}

\author{Carl M. Bender$^a$}\email{cmb@wustl.edu}
\author{Mariagiovanna Gianfreda$^{a,b}$}\email{Maria.Gianfreda@le.infn.it}

\affiliation{$^a$ Department of Physics, Washington University, St. Louis,
MO 63130, USA}
\affiliation{$^b$Institute of Industrial Science, University of Tokyo,
Komaba, Meguro, Tokyo 153-8505, Japan}


\begin{abstract}
The upside-down $-x^4$, $-x^6$, and $-x^8$ potentials with appropriate
$\cPT$-symmetric boundary conditions have real, positive, and discrete
quantum-mechanical spectra. This paper proposes a straightforward macroscopic
quantum-mechanical scattering experiment in which one can observe and measure
these bound-state energies directly.
\end{abstract}


\maketitle

In 1998 the class of $\cPT$-symmetric Hamiltonians 
\begin{equation}
H=p^2+x^2(ix)^\varepsilon
\label{e1}
\end{equation}
was introduced and it was shown numerically and perturbatively that even though
these Hamiltonians are not Hermitian, their spectra are real, positive, and
discrete if $\varepsilon>0$ \cite{r1,r2}. A rigorous proof was given in
Ref.~\cite{r3}. This spectral positivity is particularly surprising for the case
$\varepsilon=2$ because on the real axis the $-x^4$ potential is upside-down and
therefore appears to be unstable. Nevertheless, for this special value of
$\varepsilon$ there is an elementary transformation which establishes that the
Hamiltonians $H=p^2-x^4$ and $H=2p^2+4x^4-2x$ are {\it isospectral}; that is,
the eigenvalues of these two Hamiltonians are identical \cite{r4,r5,r6}. Since
the potential of the latter Hamiltonian is conventionally right-side-up, its
bound-state eigenvalues are indeed real, positive, and discrete. The numerical
values of the first three bound-state energies are
\begin{equation}
E_0=1.477150,~~E_1=6.003386,~~E_2=11.802434.
\label{e2}
\end{equation}

One might think that it is physically impossible for an upside-down potential
such as $-x^4$ to confine bound states because this potential is unbounded
below. However, the $\cPT$-symmetric eigenvalue problem for this potential must
be formulated in the complex-$x$ plane, and if we continue from real $x$ to
complex $x$, we lose the ordering principle; that is, if $x_1$ and $x_2$ are
complex, we cannot say that $x_1>x_2$ or $x_1<x_2$. Thus, the concept of a
potential $V(x)$ being unbounded below is not relevant when $x$ is complex
because the apparent instability of the $-x^4$ potential evaporates when $x$ is
complex \cite{r7}.

The objective of this paper is to show that even though the eigenvalue problem 
for the $-x^4$ potential is formulated in the complex-$x$ plane, it is actually
possible to {\it observe and measure} the bound-state energies of the $-x^4$
potential with a conceptually simple scattering experiment. The underlying idea
may be found in Ref.~\cite{r8}, where it is shown that the $-x^4$ potential
becomes reflectionless if the energy of the incident wave is equal to one of the
bound-state-energy eigenvalues. Furthermore, we will show that the bound-state
energies of the upside-down $\cPT$-symmetric $-x^6$ and $-x^8$ potentials can
also be seen and measured by using scattering experiments.

To explain the reflectionless property of upside-down $\cPT$-symmetric
potentials, we note that the bound-state solutions to the time-independent
Schr\"odinger equation for the $-x^4$ potential
\begin{equation}
-\phi''(x)-x^4\phi(x)=E\phi(x)
\label{e3}
\end{equation}
are required to vanish exponentially in an appropriate pair of Stokes wedges in
the complex-$x$ plane. To find the angular orientation of these Stokes wedges we
construct the WKB approximation to the solutions to (\ref{e3}):
\begin{equation}
\phi_{\rm WKB}(x)=C_\pm[Q(x)]^{-1/4}\exp\left[\pm i\int^x ds\sqrt{Q(s)}\right],
\label{e4}
\end{equation}
where $Q(x)=E+x^4$. Thus, for large $|x|$ the exponential component of the
asymptotic behavior of $\phi(x)$ is
\begin{equation}
\phi(x)\sim e^{\pm ix^3/3}\quad(|x|\to\infty).
\label{e5}
\end{equation}
From (\ref{e5}) we see that there are six possible Stokes wedges in the
complex-$x$ plane, each having an opening angle of $\pi/3$, inside of which
$\phi(x)$ can vanish exponentially. As explained in Ref.~\cite{r9}, we require
that $\phi(x)$ vanish in a $\cPT$-{\it symmetric pair of Stokes wedges} that are
symmetric with respect to the imaginary-$x$ axis. Substituting $x=re^{i\theta}$
in (\ref{e5}), we see that the angular range of the right Stokes wedge, which is
located adjacent to and just below the positive-real-$x$ axis, is $-\pi/3<
\theta<0$, so we must choose the {\it minus} sign in (\ref{e5}):
\begin{equation}
\phi(x)\sim e^{-ix^3/3}\quad(|x|\to\infty,~-\pi/3<{\rm arg}\,x<0).
\label{e6}
\end{equation}
The angular range of the left Stokes wedge, which is a $\cPT$ reflection of the
right Stokes wedge, is $-\pi<\theta<-2\pi/3$, so again we choose the minus sign
in (\ref{e5}):
\begin{equation}
\phi(x)\sim e^{-ix^3/3}\quad(|x|\to\infty,~-\pi<{\rm arg}\,x<-2\pi/3).
\label{e7}
\end{equation}
Equations (\ref{e6}) and (\ref{e7}) indicate that if we rotate (anticlockwise in
the right wedge and clockwise in the left wedge) to the upper edges of these
wedges, which lie along the real axis, the eigenfunctions no longer decay
exponentially and instead they represent purely oscillatory unidirectional waves
[apart from an algebraic factor of $1/|x|$ coming from the WKB approximation in
(\ref{e4})].

Similarly, for the upside-down $-x^6$ potential, which is obtained by setting
$\varepsilon=2$ in the $\cPT$-symmetric $x^4(ix)^\varepsilon$ potential, there
are two Stokes wedges of opening angle $\pi/4$ lying adjacent to and just below
the real-$x$ axis. On the real-$x$ axis the bound-state eigenfunctions, which
decay approximately like $\exp(-|x|^4)$ in these Stokes wedges, represent purely
oscillatory unidirectional waves
\begin{equation}
\phi(x)\sim e^{-ix^4/4}\quad(x\to\pm\infty)
\label{e8}
\end{equation}
(apart from an algebraic factor of $|x|^{-3/2}$). The first three bound-state
energies of the $-x^6$ potential are
\begin{equation}
E_0=1.354862,~~E_1=5.262586,~~E_2=11.234957.
\label{e9}
\end{equation}

Likewise, for the upside-down $-x^8$ potential, which is obtained by setting
$\varepsilon=2$ in the $\cPT$-symmetric $x^6(ix)^\varepsilon$ potential, there
are two Stokes wedges of opening angle $\pi/5$ lying adjacent to and just below
the real-$x$ axis. On the real-$x$ axis the bound-state eigenfunctions,
which decay approximately like $\exp(-|x|^5)$ in these Stokes wedges, represent
purely oscillatory unidirectional waves
\begin{equation}
\phi(x)\sim e^{-ix^5/5}\quad(x\to\pm\infty)
\label{e10}
\end{equation}
(apart from an algebraic factor of $|x|^{-2}$). The first three bound-state
energies of the $-x^8$ potential are
\begin{equation}
E_0=1.359859,~~E_1=5.320461,~~E_2=11.559893.
\label{e11}
\end{equation}

To summarize, {\it at the bound-state energy eigenvalues of the $\cPT$-symmetric
$-x^4$ {\rm (}and $-x^6$ and $-x^8${\rm )} potentials the eigenfunctions exhibit
asymptotically unidirectional wave propagation.} This implies that, at least in
principle, we can identify the bound-state energies by performing a scattering
experiment and looking for unidirectional incoming and outgoing waves, that
is, reflectionless scattering. However, performing an actual scattering
experiment is not trivial because these upside-down potentials increase in
strength as $|x|$ increases and they become infinite at $|x|=\infty$. Thus, in
the laboratory it is necessary to confine the potential to a finite-size box and
thus we must study the {\it cut-off} version of the potential.

Let us consider a one-dimensional scattering experiment in which the cut-off
potential $V(x)$ vanishes outside a box of size $2L$:
\begin{equation}
V(x)=\begin{cases}0 & (|x|>L), \cr -x^4 & (|x|\leq L).\end{cases}
\label{e12}
\end{equation}
The time-dependent Schr\"odinger equation that describes a scattering process
in the potential $V(x)$ is
\begin{equation}
i\psi_t(x,t)=-\psi_{xx}(x,t)+V(x)\psi(x,t).
\label{e13}
\end{equation}

In a time-independent scattering experiment we take the incident wave to be a
monochromatic plane wave of energy $E$ so that the time dependence of
$\psi(x,t)$ is simply
$$\psi(x,t)=e^{-iEt}\phi(x)$$
and (\ref{e13}) reduces to a time-independent Schr\"odinger equation of the form
in (\ref{e3}):
\begin{equation}
-\phi''(x)+V(x)\phi(x)=E\phi(x).
\label{e14}
\end{equation}
Thus, a plane wave of the form $\phi(x)=e^{iax}$ is a {\it right-going} wave if
$a>0$ and a {\it left-going} wave if $a<0$.

\begin{figure}[t!]
\begin{center}
\includegraphics[scale=0.19]{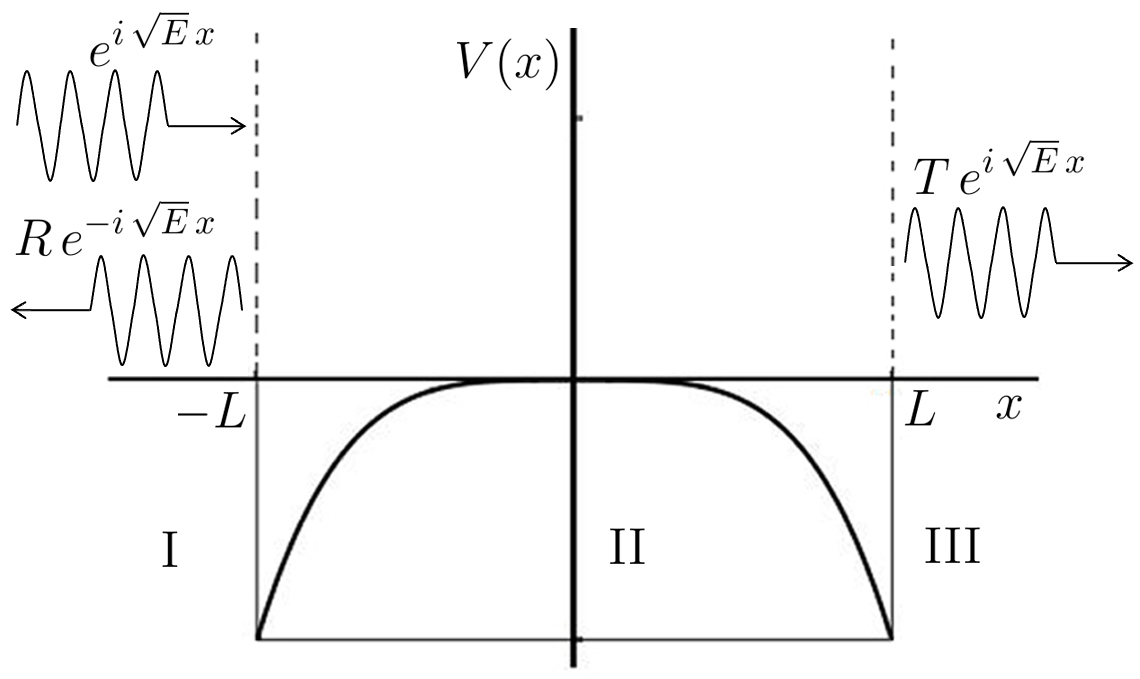}
\end{center}
\caption{Time-independent scattering experiment showing an upside-down potential
in a box of size $2L$ and the incident and reflected plane waves of energy $E$
in (\ref{e15}) and the transmitted plane wave of energy $E$ in (\ref{e16}).} 
\label{f1}
\end{figure}

There are three regions for $V(x)$ in (\ref{e12}) depending on whether $x<-L$
(Region I), $-L\leq x\leq L$ (Region II), and $x>L$ (Region III) (see
Fig.~\ref{f1}). In Region I we
assume that we have a right-going {\it incident plane wave} of unit amplitude
and energy $E$ plus a left-going {\it reflected plane wave} of amplitude $R$ and
energy $E$:
\begin{equation}
\phi_{\rm I}(x)=e^{ix\sqrt{E}}+Re^{-ix\sqrt{E}}.
\label{e15}
\end{equation}
In Region III we assume that we have a right-going {\it transmitted plane wave}
of amplitude $T$ and energy $E$:
\begin{equation}
\phi_{\rm III}(x)=Te^{ix\sqrt{E}}.
\label{e16}
\end{equation}

We then define the scaled wave function $y(x)$,
\begin{equation}
y(x)\equiv\phi(x)e^{-iL\sqrt{E}}/T,
\label{e17}
\end{equation}
and we impose the following outgoing-wave boundary conditions at $x=L$:
\begin{equation}
y_{\rm III}(L)=1,\qquad y^\prime_{\rm III}(L)= i\sqrt{E}.
\label{e18}
\end{equation}
Using these boundary conditions, we perform a numerical integration of the
time-independent Schr\"odinger equation (\ref{e14}) from $x=L$ down to $x=-L$
and compare the numerical results with the boundary conditions at $x=-L$,
which are
\begin{eqnarray}
y_{\rm I}(-L) &=& e^{-2iL\sqrt{E}}/T+R/T,\nonumber\\
y^\prime_{\rm I}(-L)&=&i\sqrt{E}e^{-2iL\sqrt{E}}/T-iR\sqrt{E}/T.
\label{e19}
\end{eqnarray}
This gives an explicit form for the ratio $|R/T|$:
\begin{equation}
|R/T|=\half|y_{\rm I}(-L)+iy'_{\rm I}(-L)/\sqrt{E}|.
\label{e20}
\end{equation}

\begin{figure}[t!]
\begin{center}
\includegraphics[scale=0.34]{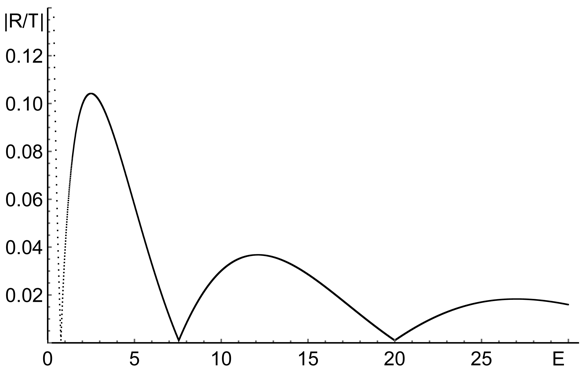} 
\end{center}
\caption{Plot of $|R/T|$ in (\ref{e20}) as a function of $E$ for $L=1$. The
vanishing of this ratio indicates that there is resonant (reflectionless)
scattering. Note that $|R/T|$ vanishes at $E=0.76$, $7.55$, and $19.99$.
However, comparing this figure with Figs.~\ref{f3} and \ref{f4}, we see that the
energies at which resonant (reflectionless) scattering occurs depend on
$L$. Furthermore, these zeros do not agree with the eigenvalues in (\ref{e2}).}
\label{f2}
\end{figure}

\begin{figure}[h!]
\begin{center}
\includegraphics[scale=0.34]{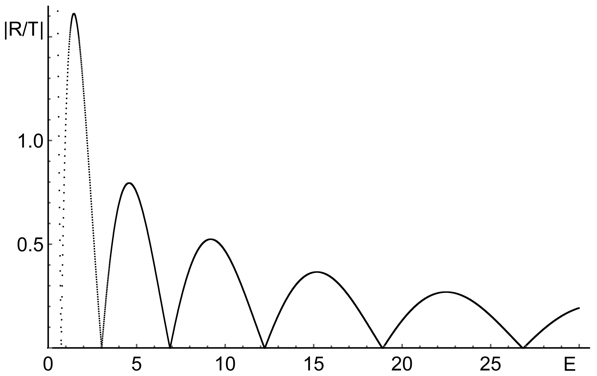} 
\end{center}
\caption{Plot of the ratio $|R/T|$ in (\ref{e20}) as a function of $E$ for $L=
2$. The zeros at which resonant scattering occurs are now located at $E=0.74$,
$3.02$, $6.88$, $12.23$, $18.89$, and $26.83$ and they have moved from their
positions in Fig.~\ref{f2}. This shows that the resonant scattering energies
depend on $L$.}
\label{f3}
\end{figure}

\begin{figure}[h!]
\begin{center}
\includegraphics[scale=0.34]{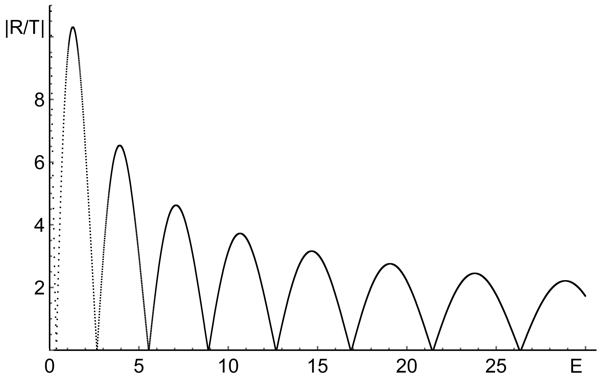} 
\end{center}
\caption{Plot of $|R/T|$ in (\ref{e20}) as a function of $E$ for $L=5$. The
zeros are now located at $E=0.38$, $2.75$, $5.55$, $8.89$, $12.68$, $16.88$,
$21.44$, and $26.34$.}
\label{f4}
\end{figure}

In Figs.~\ref{f2}--\ref{f4} we plot the computed numerical ratio $|R/T|$ in
(\ref{e20}) for $L=1$, $2$, and $5$. Observe that the zeros of $|R/T|$ depend on
the value of $L$. Furthermore, these zeros are clearly not related to the
eigenvalues in (\ref{e2}). Therefore, if in this experiment the incident,
reflected, and transmitted plane waves have energy $E$, we cannot measure the
$\cPT$-symmetric bound-state energies. Evidently, this way of performing the
scattering experiment is not successful because the zeros of $|R/T|$ shown in
Figs.~\ref{f2}--\ref{f4} are merely the conventional resonant scattering
energies of the potential $V(x)$ at which the potential becomes reflectionless.
These resonant energies {\it depend on the size $2L$ of the box}.

There {\it is} a way to perform a time-independent scattering experiment in
which we can observe the energy levels in (\ref{e2}). Using plane waves only, we
take the energies of the incident, reflected, and transmitted waves in Regions I
and III to {\it depend on} $L$. To understand why this works, let us examine
more carefully the WKB approximation to the solution to the time-independent
Schr\"odinger equation (\ref{e3}). Because the WKB approximation to $\phi(x)$
has the form in (\ref{e4}), we choose the following $L$-dependent boundary
conditions on the positive-$x$ axis at $x=L$:
\begin{equation}
\phi_{\rm III}(L)=1,~~\phi^\prime_{\rm III}(L)=i\sqrt{Q(L)}-\quarter Q'(L)/Q(L).
\label{e21}
\end{equation}
To obtain these simple boundary conditions we have set $C_-=0$ in (\ref{e4})
(because in Region III there are only right-going waves) and we have chosen
$C_+$ so that
$$\phi_{\rm WKB}(x)=\left[\frac{Q(L)}{Q(x)}\right]^{1/4}\exp\left[
i\int_L^x ds\sqrt{Q(s)}\right].$$
Of course, to justify the using a WKB approximation we must assume that $L$ is
large and that $x$ is near $L$.

Then, on the negative-$x$ axis we take a WKB approximation of the form
\begin{eqnarray}
\phi_{\rm WKB}(x)&=&\left[\frac{Q(L)}{Q(x)}\right]^{1/4}\left(
D_-\exp\left[-i\int_{-L}^x ds\sqrt{Q(s)}\right]\right.\nonumber\\
&&\quad+\left.D_+\exp\left[i\int_{-L}^x ds\sqrt{Q(s)}\right]\right),\nonumber
\end{eqnarray}
where we assume that $x$ is near $-L$. (Note here that $D_+$ is the
coefficient of the right-going incident wave and $D_-$ is the coefficient of
the left-going reflected wave.) From this formula we calculate
$$\phi_{\rm I}(-L)=D_-+D_+$$
and
\begin{eqnarray}
\phi_{\rm
I}'(-L)&=&D_-\left[-i\sqrt{Q(L)}-\quarter Q'(L)/Q(L)\right]\nonumber\\
&&\quad+D_+\left[i\sqrt{Q(L)}-\quarter Q'(L)/Q(L)\right].\nonumber
\end{eqnarray}
Thus, 
\begin{equation}
\left|D_-\right|=\frac{1}{2}\left|\left[i-
\frac{Q'(L)}{4Q^{3/2}(L)}\right]\phi(-L)-\frac{\phi'(-L)}{\sqrt{Q(L)}}\right|.
\label{e22}
\end{equation}

\begin{figure}[t!]
\begin{center}
\includegraphics[scale=0.34]{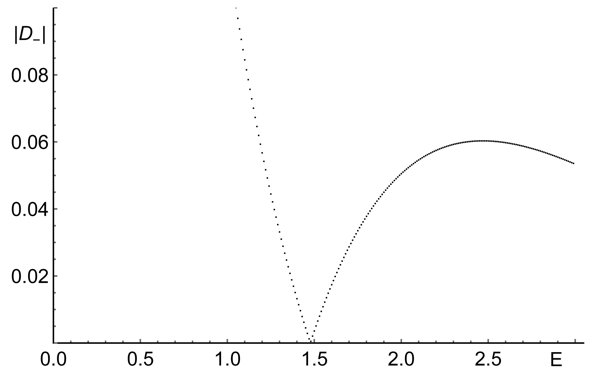}
\end{center}
\caption{Plot of $|D_-|$ in (\ref{e22}) as a function of $E$ for $L=5$.
There is a zero at $E=1.475$, which is in very good agreement with the exact
value of the ground-state energy $1.477$ in (\ref{e2}).}
\label{f5}
\end{figure}

\begin{figure}[h!]
\begin{center}
\includegraphics[scale=0.34]{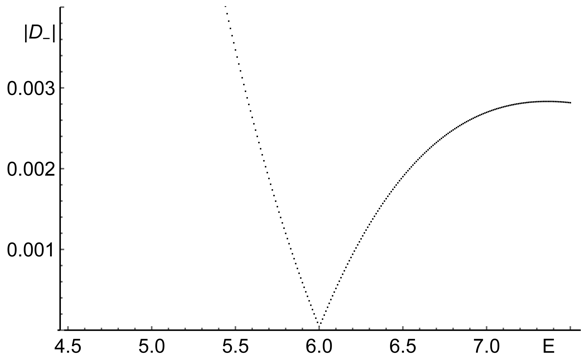} 
\end{center}
\caption{Plot of $|D_-|$ in (\ref{e22}) as a function of $E$ for $L=5$. The
zero lies at $E=6.005$, which is an accurate approximation to the exact value
of the first energy level $6.003$ in (\ref{e2}).}
\label{f6}
\end{figure}

\begin{figure}[h!]
\begin{center}
\includegraphics[scale=0.34]{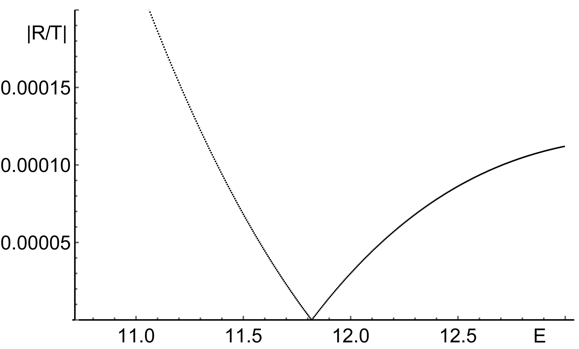} 
\end{center}
\caption{Plot of $|D_-|$ in (\ref{e22}) as a function of $E$ for $L=7$.
The zero at $11.820$ is an accurate approximation to the
exact value of the second energy level $11.802$ in (\ref{e2}).
If we take $L=5$, the zero is at $E=11.700$, which is not quite as accurate.}
\label{f7}
\end{figure}

We now integrate the Schr\"odinger equation (\ref{e3}) from $L$ to $-L$ using
the boundary conditions (\ref{e21}) and calculate $\phi_{\rm I}(-L)$ and
$\phi_{\rm I}'(-L)$ numerically. Then from (\ref{e22}) we calculate the
coefficient $D_-$ of the reflected wave in Region I. In Figs.~\ref{f5}--\ref{f7}
we plot $|D_-|$. Note that $|D_-|$ vanishes at the bound-state energies of the
$-x^4$ potential in (\ref{e2}).

\begin{figure}[h!]
\begin{center}
\includegraphics[scale=0.34]{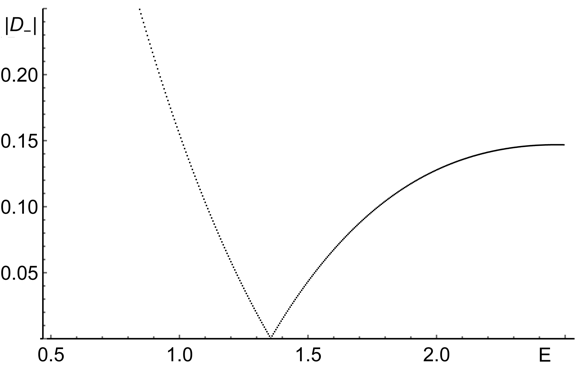} 
\end{center}
\caption{Plot of $|D_-|$ for the $-x^6$ potential for $L=5$. There is a zero
at $1.355$, which agrees with $E_0=1.355$ in (\ref{e9}).}
\label{f8}
\end{figure}

\begin{figure}[h!]
\begin{center}
\includegraphics[scale=0.34]{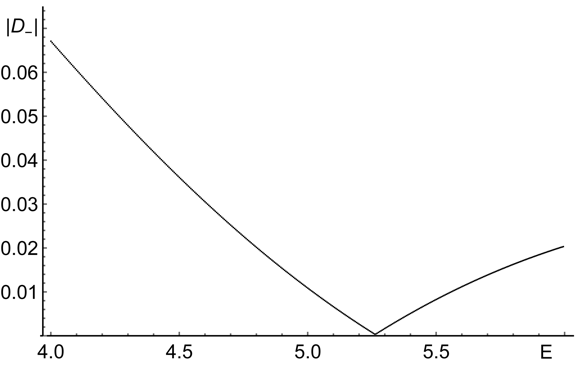}
\end{center}
\caption{Plot of $|D_-|$ for the potential $-x^6$ for $L=5$. The zero 
at $5.270$ agrees with the exact value of $E_1=5.263$ in (\ref{e9}).}
\label{f9}
\end{figure}

\begin{figure}[h!]
\begin{center}
\includegraphics[scale=0.34]{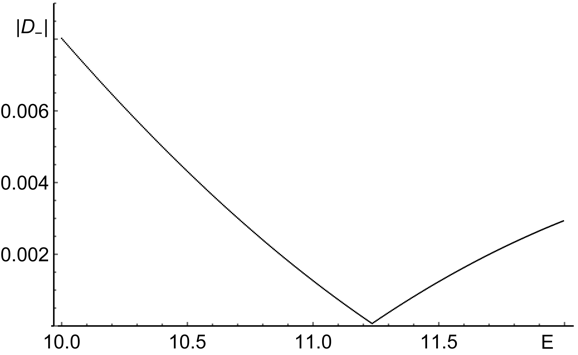}
\end{center}
\caption{Plot of $|D_-|$ for the potential $-x^6$ for $L=5$. The zero
at $11.240$ agrees with the exact value $E_2=11.235$ in (\ref{e9}).}
\label{f10}
\end{figure}

\begin{figure}[h!]
\begin{center}
\includegraphics[scale=0.34]{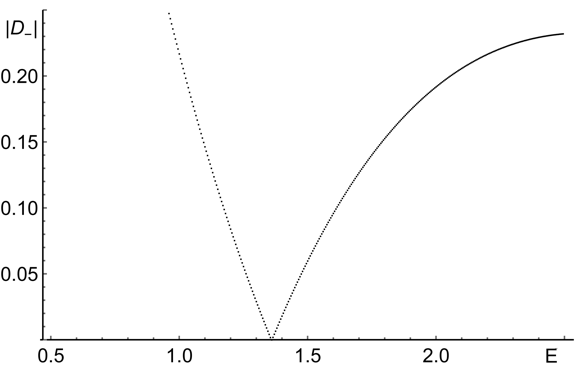}
\end{center}
\caption{Plot of $|D_-|$ for the potential $-x^8$ for $L=5$. The zero
at $1.360$ compares well with $E_0=1.359$ in (\ref{e11}).}
\label{f11}
\end{figure}

\begin{figure}[h!]
\begin{center}
\includegraphics[scale=0.34]{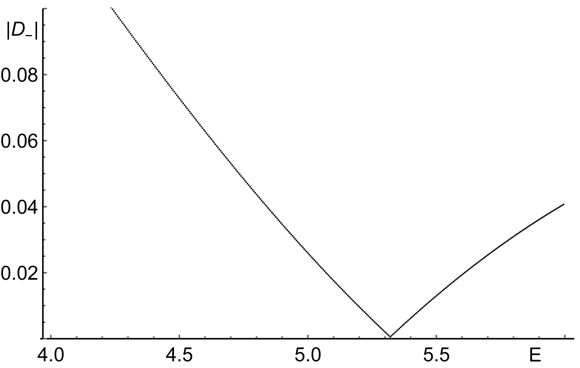}
\end{center}
\caption{Plot of $|D_-|$ for the potential $-x^8$ for $L=5$. The zero
at $5.320$ compares well with $E_1=5.320$ in (\ref{e11}).}
\label{f12}
\end{figure}

\begin{figure}[b!]
\begin{center}
\includegraphics[scale=0.34]{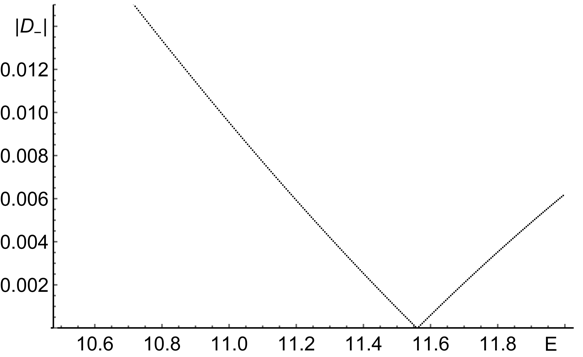}
\end{center}
\caption{Plot of $|D_-|$ for the potential $-x^8$ for $L=5$. The zero
at $11.560$ compares well with $E_2=11.560$ in (\ref{e11}).}
\label{f13}
\end{figure}

Our results for the $-x^4$ potential immediately generalize to other
upside-down potentials. For a $-x^6$ potential we take $Q(x)=E+x^6$. The
analogs of Figs.~\ref{f5}--\ref{f7} are Figs.~\ref{f8}--\ref{f10}. For a
$-x^8$ potential we take $Q(x)=E+x^8$ and the analogs of
Figs.~\ref{f5}--\ref{f7} are Figs.~\ref{f11}--\ref{f13}.

To conclude, we emphasize that the spectrum of an upside-down $\cPT$-symmetric
potential is real, positive, and discrete. This remarkable and rigorous
mathematical result reflects the nature of boundary-value problems in the
complex domain. However, in this paper we have shown that it is possible to
observe these eigenvalues in the laboratory by using a simple macroscopic
time-independent quantum-mechanical scattering experiment. (To produce an
upside-down $-x^4$ potential in a box one might superpose a cosine potential
associated with the Josephson effect with a parabolic potential.) Many
experimental studies of classical $\cPT$-symmetric systems have been done
\cite{r10,r11,r12,r13,r14,r15,r16,r17,r18} but these experiments have focused on
the classical $\cPT$ phase transition. The experiment proposed in this paper
would be the first experimental examination of the bound-state spectrum of a
quantum-mechanical $\cPT$-symmetric system.

\acknowledgments
We thank D.~Hook for precise numerical calculations of bound-state energies
and K.~Murch for discussions about experimental possibilities.

\end{document}